\documentclass[12pt]{article}

\begin{document}

\author{Gyula Fodor \\
KFKI Research Institute for Particle and Nuclear Physics,\\
H-1525, Budapest 114, P. O. B. 49, Hungary\\
Department of Physics, Waseda University, 3-4-1 Okubo,\\
Shinjuku, Tokyo 169-8555, Japan\\
email: gfodor@rmki.kfki.hu}
\title{Generating spherically symmetric static perfect fluid solutions}
\maketitle

\begin{abstract}
By a choice of new variables the pressure isotropy condition for 
spherically symmetric static perfect fluid spacetimes can be made a 
quadratic algebraic equation in one of the two functions appearing in it.
Using the other variable as a generating function, the pressure and the 
density of the fluid can be expressed algebraically by the function and 
its derivatives. One of the functions in the metric can also be expressed 
similarly, but to obtain the other function, related to the redshift 
factor, one has to perform an integral. Conditions on the generating 
function ensuring regularity and physicality near the center are 
investiagted. Two everywhere physically well behaving example solutions 
are generated, one representing a compact fluid body with a zero pressure 
surface, the other an infinite sphere.
\end{abstract}

\section{Introduction}

The aim of this paper is to discuss an algorithm that can be used to
generate any number of physically realistic pressure and density profiles
for spherical perfect fluid distributions without calculating integrals. Our
key step is the transformation of the field equation into a form algebraic
in one variable. A similar equation was already written down in the paper of
Burlankov\cite{burl}. However, it was immediately transformed to isotropic
coordinates, thereby obtaining a generating formalism similar to that of
Glass and Goldman\cite{gago}\cite{gold}, where even the calculation of
pressure and density requires integrals.

Because of their inability to support shear stresses static perfect fluid
configurations are expected to be spherically symmetric. Theorems showing
this have been proved under fairly general conditions\cite{besi}\cite{lima}.
In the static spherically symmetric case the only relation one gets from the
Einstein equations is the pressure isotropy condition. This condition
together with a fluid equation of state and a central density value can be
used to determine a spherically symmetric regular matter distribution, which
either connects to an exterior vacuum region through a zero pressure
hypersurface or extends to the whole spacetime. The uniqueness of this
spacetime has been proved by Rendall and Schmidt\cite{resc} for monotonic
equation of states. Regularity of the metric can be shown even without the
monotonicity assumption\cite{bare}.

Since the addition of even the simplest equation of state makes the task of
finding an exact solution extremely difficult, the usual method to proceed
is to look for some solution of the pressure isotropy condition under some
mathematical assumptions and hope that the resulting equation of state will
be simple and physical. The most common way to choose the radial coordinate $%
r$ is by setting the surface of the isometry spheres to $4\pi r^{2}.$ Then
by appropriately choosing the form of the two functions describing the
metric, the pressure isotropy condition is a first order linear equation in
one of the two variables. As already was pointed out by Wyman\cite{wyman},
for arbitrary choice of the other function this equation can always be
solved by quadratures. In the other variable the pressure isotropy condition
is either a second order linear differential equation or a first order
Riccati equation, depending on the exact choice of the function. Considering
this relative mathematical simplicity, it is not surprising that more than
hundred static perfect fluid exact solutions have already appeared in the
literature\cite{dela}. Unfortunately a large part of these solutions are
unphysical in several aspects. Many do not have a regular center, others
have no positive pressure and density, violate the dominant energy
condition, or have unphysical sound speed.

Even though the field equation is in principle solvable by quadratures, the
resulting integrals can be calculated in terms of elementary functions only
in some exceptional cases. The imposition of physicality near the center
makes the possible generating functions even more complicated and the chance
of being able to calculate the integrals even less. There are some
techniques that can be used to generate new solutions from known ones
published in the literature\cite{hein}\cite{whit}, but they also require
calculation of integrals or solution of differential equations. Using
isotropic coordinates, Kuchowicz\cite{kuch} writes the pressure isotropy
condition into a form which is algebraic in one of its two variables. Glass
and Goldman\cite{gago}\cite{gold} introduce pressure and density related
variables, also in isotropic coordinates, to obtain a similar algebraic
equation. 
They use one of the two variables as a generating function, and obtain 
formulae for the pressure, the density and the metric functions involving 
integrals of functional expressions of the generating function and its 
derivatives.
They also give conditions on the generating function which ensure that the 
resulting metric is physically realistic near the origin. 
In the present paper we follow a
similar approach in area coordinates $r.$ The main advantage of our proposed
method is that it does not require the calculation of any integrals for the
expression of the pressure, the density and one of the metric functions. A
similar equation to our main equation has been already calculated by
Burlankov\cite{burl}, but a transformed form of it was used there to
generate solutions.

In Section \ref{fieq} we transform the pressure isotropy condition into a
quadratic algebraic equation in one of its variables and propose to use the
other variable as a generating function. 
The pressure, the density and one of the
functions in the metric are expressed as algebraic expressions of the
generating function and its first and second derivatives. The other
function, determining the redshift factor, appears only in differentiated
form in the field equation, and consequently only its derivative can be
given in an algebraic form. The functional form of the generating function
is calculated for some known solutions in Section \ref{known}. A power
series expansion near the center is used in Section \ref{secexp} in order to
establish necessary conditions on the generating function to make the center
regular, the density and pressure positive, having a local maximum, and
satisfying the dominant energy condition near the center. The metric induced
by the simplest physically realistic polynomial generating function is
calculated in Section \ref{gensol}. The solution represents a fluid sphere
with a regular center and a zero pressure surface. It has physical density
and pressure and casual sound speed for some choice of the parameters. In
Section \ref{gensol2}, a second simple choice of the generating function is
used to calculate a further solution, for which the matter extends to
infinity.

\section{Field equations}

\label{fieq}The metric $g_{\mu \nu }$ of a general stationary spherically
symmetric configuration can be written in area coordinates as 
\begin{equation}
ds^{2}=-e^{2\nu }dt^{2}+\frac{1}{B}dr^{2}+r^{2}\left( d\vartheta ^{2}+\sin
^{2}\vartheta d\varphi ^{2}\right)  \label{ds1}
\end{equation}
where $\nu $ and $B$ are functions of the radial coordinate $r.$ We assume
that the spacetime region is filled with perfect fluid, and the fluid
velocity vector $u^{\mu }$ is proportional to $\partial /\partial t.$ Since
solutions of the Einstein equations with nonzero cosmological constant can
be interpreted as solutions with negative pressure or density contributions,
we do not include a cosmological term explicitly in our equations.
Calculating the Einstein equations $G_{\mu \nu }=8\pi T_{\mu \nu },$ we
obtain the energy density, the radial and the angular directional pressures
as 
\begin{eqnarray}
\mu &=&\frac{1}{8\pi r^{2}}\left( 1-B-rB^{\prime }\right) \ ,  \label{mu0} \\
p_{r} &=&\frac{1}{8\pi r^{2}}\left( 2rB\nu ^{\prime }+B-1\right) \ ,
\label{pr0} \\
p_{\vartheta } &=&\frac{1}{16\pi r}\left[ 2rB\nu ^{\prime \prime }+2rB(\nu
^{\prime })^{2}+r\nu ^{\prime }B^{\prime }+2B\nu ^{\prime }+B^{\prime }%
\right] \ ,
\end{eqnarray}
where the prime denotes derivatives with respect to the radial coordinate $%
r. $ The only field equation one gets is the pressure isotropy condition $%
p\equiv p_{r}=p_{\vartheta ,}$ which takes the form 
\begin{equation}
r\left( r\nu ^{\prime }+1\right) B^{\prime }+\left[ 2r^{2}\nu ^{\prime
\prime }+2r^{2}(\nu ^{\prime })^{2}-2r\nu ^{\prime }-2\right] B+2=0\ .
\label{ein0}
\end{equation}
This equation together with an equation of state $p=p(\mu )$ and a central
density $\mu _{c}$ determine a unique spherically symmetric perfect fluid
configuration, at least in the $\frac{dp}{d\mu }>0$ case\cite{resc}. In this
paper, however, we try to find the general solution of the pressure isotropy
condition $p_{r}=p_{\vartheta }$ and leave the task of interpreting the
resulting equation of states as a second step. Equation (\ref{ein0}) is
first order and linear in $B,$ and consequently, for any given function $\nu
,$ the general solution for $B$ can be given in terms of integrals\cite
{wyman}. We instead proceed by transforming (\ref{ein0}) into a simpler form.

Because of the freedom in constant rescaling of the time coordinate $t,$
only the derivatives of the function $\nu $ appears in the expressions for $%
\mu ,$ $p_{r}$ and $p_{\vartheta }.$ This would suggest to introduce $\nu
^{\prime }$ as a new variable. However, considering the coefficient of $%
B^{\prime }$ in (\ref{ein0}), it turns out to be more useful to introduce a
new function as 
\begin{equation}
\beta =r\nu ^{\prime }+1\ .  \label{beta}
\end{equation}
Then the field equation takes the simple form 
\begin{equation}
r\beta B^{\prime }+2rB\beta ^{\prime }+2\beta ^{2}B-8\beta B+4B+2=0\ ,
\label{ein1}
\end{equation}
and the radial pressure becomes 
\begin{equation}
p_{r}=\frac{1}{8\pi r^{2}}\left( 2\beta B-B-1\right) \ .  \label{pr1}
\end{equation}

Introducing a further new variable 
\begin{equation}
\alpha =\beta ^{2}B\ ,  \label{alpha}
\end{equation}
equation (\ref{ein1}) becomes a second order algebraic equation in $\beta ,$%
\begin{equation}
2(\alpha +1)\beta ^{2}+(r\alpha ^{\prime }-8\alpha )\beta +4\alpha =0\ .
\label{ein2}
\end{equation}

If instead of $\alpha $ one introduces its square root, $z=\sqrt{\alpha }$ 
as a new function, and denote $\sqrt{B}=b,$ then it is possible to get from 
(\ref{ein1}) an equation algebraic in $b,$ which is already given in the
paper of Burlankov\cite{burl}, 
\begin{equation}
2b^{2}+\left( rz^{\prime }-4z\right) b+z^{2}+1=0\ .  \label{einb}
\end{equation}
Although at first sight this equation appears as a more comfortable form
directly giving the metric function $b=\sqrt{B}$ from the generating
function $z=\sqrt{\alpha },$ it has serious disadvantages. First of all, for
most known perfect fluid solutions $B$ and $\alpha $ are not a square, and
the appearance of the further square roots makes the calculations even more
cumbersome. The more serious second problem is that the simplest polynomial
choices for $z$ do not appear to give appropriate results. For example, the
quadratic equation (\ref{einb}) has no real solution for $b$ when choosing $%
z=1-ar^{2}$ with $a$ constant, while $\alpha =1-ar^{2}$ gives the Einstein
static universe, as we will see in Section \ref{known}. In general,
introducing any functional form of $\alpha ,$ such as $\alpha ^{2},$ $\frac{1%
}{\alpha }$ or $e^{\alpha },$ as a new variable would give a different
algebraic equation in the other variable. This makes our formalism
non-unique in one hand, but gives opportunities to generate even more
variety of new solutions on the other hand.

A further disadvantage of the equation (\ref{einb}) compared to (\ref{ein2})
that it is nonlinear in both variables. Since equation (\ref{ein2}) is
linear in $\alpha ,$ if the function $\nu $ and consequently $\beta =r\nu
^{\prime }+1$ is given, one can express its general solution for $\alpha $
using integrals,
\begin{eqnarray}
\alpha &=&e^{-I_{\beta }}\left( C-2\int \frac{\beta }{r}e^{I_{\beta
}}dr\right) \;,\;\; \\
I_{\beta } &=&\int \frac{2}{r\beta }\left( \beta ^{2}-4\beta +2\right) dr\ ,
\end{eqnarray}
where $C$ is some constant. Unfortunately, there is no guarantee, that for
those $\nu $ for which the integrals can be given in terms of elementary
functions, the resulting equations of states will be also physically
reasonable. Instead of trying to investigate this further we focus on the
alternative approach, considering the function $\alpha $ as the basic
quantity.

For any function $\alpha $ for which 
\begin{equation}
(8\alpha -r\alpha ^{\prime })^{2}>32\alpha (\alpha +1)  \label{alco}
\end{equation}
the quadratic equation (\ref{ein2}) has two solutions for the function $%
\beta .$ We denote them by $\beta _{+}$ and $\beta _{-},$ 
\begin{equation}
\beta _{\pm }=\frac{1}{4(\alpha +1)}\left[ 8\alpha -r\alpha ^{\prime }\pm 
\sqrt{(8\alpha -r\alpha ^{\prime })^{2}-32\alpha (\alpha +1)}\right] \ .
\label{besq}
\end{equation}
The condition (\ref{alco}) will turn out to be not particularly restrictive,
since as we will see in Section \ref{secexp}, at a regular center $\alpha
\longrightarrow 1$ and then (\ref{alco}) holds as an equality at $r=0.$
Furthermore, we will see that positivity of the fluid density will ensure
the existence of the square root near the center. Actually, it will also
turn out that the root $\beta _{-}$ always belong to non-positive densities,
and hence it is unphysical. For actual calculations it may be convenient to
use the identity 
\begin{equation}
\beta _{+}\beta _{-}=\frac{2\alpha }{\alpha +1}
\end{equation}
to eliminate square roots from the denominator.

Using (\ref{alpha}) and (\ref{beta}), the metric functions $B_{+}$ and $\nu
_{+}$ belonging to $\beta _{+},$ and $B_{-}$ and $\nu _{-}$ belonging to $%
\beta _{-},$ can be calculated as 
\begin{eqnarray}
B_{\pm } &=&\frac{\alpha }{\beta _{\pm }^{\ 2}}=\frac{(\alpha +1)^{2}}{%
4\alpha }\ \beta _{\mp }^{\ 2},  \label{eqB} \\
\nu _{\pm}&=&\int\limits_{0}^{r}\frac{1}{r}(\beta_{\pm}-1)dr+C_{\pm}\ ,
\label{eqA}
\end{eqnarray}
where $C_{+}$ and $C_{-}$ are constants determining the scaling of the time
coordinate $t.$ The integral (\ref{eqA}) generally cannot be given in terms
of elementary functions. However, as can be seen from (\ref{mu0}) and (\ref
{pr1}), the physically important pressure and density can be calculated
without performing integrals. Denoting the pressure and density belonging to 
$\beta _{+}$ by $p_{+}$ and $\mu _{+},$ and those belonging to $\beta _{-}$
by $p_{-}$ and $\mu _{-}$, 
\begin{eqnarray}
p_{\pm } &=&\frac{1}{8\pi r^{2}}\left( 2\beta _{\pm }B_{\pm }-B_{\pm
}-1\right) \ ,  \label{pp2} \\
\mu _{\pm } &=&\frac{1}{8\pi r^{2}}\left( 1-B_{\pm }-rB_{\pm }^{\prime
}\right) \ .  \label{mu2}
\end{eqnarray}

The form of the equation of state is extremely complicated in general, and
possibly cannot even be put into an $r$ independent $\mu=\mu(p)$ or
$f(\mu,p)=0$ form using elementary functions. 
However, the important point is that, in principle, all static spherically 
symmetric perfect fluid solutions, with all possible equations of states, 
could be given by suitably choosing the function $\alpha.$ 
Indeed, for any known solution, first $\beta,$ then $\alpha$ can
be calculated in terms of the metric functions $\nu$ and $B,$ using the
equations (\ref{beta}) and (\ref{alpha}).

\section{Some known solutions}

\label{known}In this section we look at the form of some known exact
solutions in our formalism. Since the field equation (\ref{ein2}) has in
general two roots, most of these metrics are paired with a perfect fluid
counterpart solution.

The simplest example is the Minkowski spacetime with $\alpha =1.$ This is
the only constant $\alpha $ spacetime with a regular center. Then the
quadratic equation (\ref{ein2}) has only one solution, $\beta =1.$

Next we consider the vacuum Schwarzschild spacetime with mass parameter $m.$
Then $e^{2\nu }=B=1-2m/r,$ and from (\ref{beta}) and (\ref{alpha}) 
\begin{eqnarray}
\beta &=&\frac{r-m}{r-2m}\ ,  \label{besch} \\
\alpha &=&\frac{(m-r)^{2}}{r(r-2m)}=-\frac{m}{2r}+\frac{3}{4}-\frac{r}{8m}-%
\frac{r^{2}}{16m^{2}}+O(r^{3})\ .
\end{eqnarray}
However, given this form of $\alpha ,$ (\ref{besch}) is only one of the
solutions of (\ref{ein2}). The other solution, corresponding to the negative
sign in (\ref{besq}) is 
\begin{equation}
\beta _{-}=2\frac{(r-m)(r-2m)}{m^{2}-4mr+2r^{2}}\ .
\end{equation}
This belongs to a perfect fluid spacetime with pressure and density 
\begin{eqnarray}
p_{-} &=&\frac{m^{3}(7m-4r)}{32\pi r^{3}(r-2m)^{3}}\ , \\
\mu _{-} &=&\frac{m^{2}(2r-3m)(10r-17m)}{32\pi r^{2}(r-2m)^{4}}\ .
\end{eqnarray}
The solution has a complicated equation of state and does not have a
regular center at $r=0.$

The next simplest solution with a regular center is the Einstein static
universe with constant $\nu ,$ with $\beta =1$ and $\alpha =B=1-ar^{2},$
where $a$ is some positive constant. The other root of (\ref{ein2}) is 
\begin{equation}
\beta _{-}=2\frac{1-ar^{2}}{2-ar^{2}}\ .
\end{equation}
Using (\ref{eqB}) and(\ref{eqA}), the functions in the metric (\ref{ds1})
turn out to be 
\begin{equation}
e^{2\nu _{-}}=C(2-ar^{2})\;,\;\;B_{-}=\frac{(2-ar^{2})^{2}}{4(1-ar^{2})}\ .
\end{equation}
The pressure 
\begin{equation}
p_{-}=\frac{a(3ar^{2}-4)}{32\pi (1-ar^{2})}
\end{equation}
can be positive for negative $a,$ but the density 
\begin{equation}
\mu _{-}=\frac{a^{2}r^{2}(3ar^{2}-5)}{32\pi (1-ar^{2})^{2}}
\end{equation}
is always negative near the center and vanishes for $r=0.$ Although this
solution is unphysical in the zero cosmological constant case, it has a
regular center, and the fluid obeys a remarkably simple equation of state 
\begin{equation}
8\pi a\mu _{-}=-(a+8\pi p_{-})(3a+64\pi p_{-})\ .
\end{equation}

The interior Schwarzschild solution is described by the metric components 
\begin{equation}
e^{2\nu }=b^{2}\left( a-\sqrt{1-\frac{r^{2}}{R^{2}}}\right) ^{2}\;,\;\;B=1-%
\frac{r^{2}}{R^{2}}\ ,
\end{equation}
where $a,$ $b$ and $R$ are constants. The pressure is positive if $1<a<3.$
This metric can be obtained in our formalism by choosing 
\begin{eqnarray}
\alpha &=&\left[ \sqrt{1-\frac{r^{2}}{R^{2}}}+\frac{r^{2}}{R^{2}\left( a-%
\sqrt{1-\frac{r^{2}}{R^{2}}}\right) }\right] ^{2} \\
&=&1+\frac{3-a}{R^{2}(a-1)}r^{2}-\frac{r^{4}}{R^{4}(a-1)}-\frac{a^{2}-3a+4}{%
4R^{6}(a-1)^{3}}r^{6}+O(r^{8})  \nonumber
\end{eqnarray}
and taking the root 
\begin{equation}
\beta _{+}=1+\frac{r^{2}}{r^{2}-R^{2}+aR^{2}\sqrt{1-\frac{r^{2}}{R^{2}}}}
\end{equation}
of the quadratic equation (\ref{ein2}). The other solution $\beta _{-}$
gives another perfect fluid metric with a regular center and a complicated
equation of state. Unfortunately, the matter density $\mu _{-}$ turns out to
be negative near the center when the pressure is positive. We will see in
the next section that this is a general property. If one solution of (\ref
{ein2}) is physically well behaving, then the other belongs to negative
matter densities.

\section{Expansion around a regular center}

\label{secexp}In order for the metric (\ref{ds1}) to possess a regular
center the functions $\nu $ and $B$ must have regular limits at $r=0.$ To
obtain the standard area per radius ratio for small spheres the limit of $B$
must be $1.$ Further restrictions may come from differentiability conditions
in a coordinate system which is regular at the center, and also from the
regularity of the pressure and density. It can be seen from (\ref{beta}) and
(\ref{alpha}) that the limit of the functions $\beta $ and $\alpha $ at $r=0$
must be also $1.$

We take the expansion of the function $\alpha $ in the form 
\begin{equation}
\alpha =\sum_{i=0}^{\infty }\alpha _{i}r^{i}\ ,  \label{aex}
\end{equation}
where $\alpha _{i}$ are constant expansion coefficients, and $\alpha _{0}=1$%
. If we assumed that $\nu $ and $B,$ and consequently $\alpha $ and $\beta ,$
are analytic at the center in a rectangular coordinate system, then because
of the spherical symmetry, all their odd expansion coefficients would have
to vanish, i. e. $\alpha _{i}$ would have to be zero for odd $i.$ We proceed
without this assumption now, and examine the first few expansion
coefficients one by one.

If $\alpha _{1}$ is nonzero, calculating $\beta $ from (\ref{besq}), $B$
from (\ref{eqB}), $p$ and $\mu $ from (\ref{pp2}) and (\ref{mu2}), the
leading term in the pressures $p_{+}$ and $p_{-}$ turns out to be
proportional to $\frac{\alpha _{1}}{r},$ and the leading term in the
densities $\mu _{+}$ and $\mu _{-}$ is proportional to $\frac{\sqrt{\alpha
_{1}}}{r^{3/2}}.$ If $\alpha _{1}=0$ then the pressure is regular but the
densities are proportional to $\frac{\sqrt{\alpha _{3}}}{\sqrt{r}}.$ This
shows that $\alpha _{1}$ and $\alpha _{3}$ must be zero for regular matter
distributions. No such restriction follows for $\alpha _{5}$ and higher odd
index coefficients.

When $\alpha _{1}=\alpha _{3}=0$ then the expression under the square root
in (\ref{besq}) has the expansion 
\begin{eqnarray}
(8\alpha -r\alpha ^{\prime })^{2}-32\alpha (\alpha +1) &=&4(\alpha
_{2}^{2}-8\alpha _{4})r^{4} \\
&&\hspace*{-2.5cm}-48\alpha _{5}r^{5}-16(\alpha _{2}\alpha _{4}+4\alpha
_{6})r^{6}+O(r^{7})\ .  \nonumber
\end{eqnarray}
In order for $\beta $ to exist, this must be non-negative, which holds near 
the center only if $8\alpha_{4}\leq\alpha_{2}^{2}.$ There are two different
spacetimes belonging to a given $\alpha.$ One belongs to the positive sign
in (\ref{besq}), and the other to the negative sign. The expansions of the
pressures belonging to $\beta _{+}$ and $\beta _{-}$ are 
\begin{eqnarray}
8\pi p_{+} &=&\alpha _{2}-\frac{1}{8}\left( \alpha _{2}^{2}+\alpha _{2}\sqrt{%
\alpha _{2}^{2}-8\alpha _{4}}-12\alpha _{4}\right) r^{2}  \nonumber \\
&&+\frac{\alpha _{5}}{4}\left( \frac{3\alpha _{2}}{\sqrt{\alpha
_{2}^{2}-8\alpha _{4}}}+7\right) r^{3}+O(r^{4})\ ,  \label{ppex} \\
8\pi p_{-} &=&\alpha _{2}-\frac{1}{8}\left( \alpha _{2}^{2}-\alpha _{2}\sqrt{%
\alpha _{2}^{2}-8\alpha _{4}}-12\alpha _{4}\right) r^{2}  \nonumber \\
&&-\frac{\alpha _{5}}{4}\left( \frac{3\alpha _{2}}{\sqrt{\alpha
_{2}^{2}-8\alpha _{4}}}-7\right) r^{3}+O(r^{4})\ .
\end{eqnarray}
These readily show that the pressure can be positive only if $\alpha _{2}>0.$
The expansions of the fluid densities are 
\begin{eqnarray}
8\pi \mu _{+} &=&\frac{3}{2}\left( \sqrt{\alpha _{2}^{2}-8\alpha _{4}}%
-\alpha _{2}\right) -\frac{12\alpha _{5}}{\sqrt{\alpha _{2}^{2}-8\alpha _{4}}%
}r \\
&&-\frac{5}{8}\left[ \alpha _{2}^{2}-4\alpha _{4}+\frac{\alpha
_{2}^{3}+32\alpha _{6}}{\sqrt{\alpha _{2}^{2}-8\alpha _{4}}}+\frac{72\alpha
_{5}^{2}}{\left( \alpha _{2}^{2}-8\alpha _{4}\right) ^{\frac{3}{2}}}\right]
r^{2}+O(r^{3})\ ,  \nonumber \\
8\pi \mu _{-} &=&-\frac{3}{2}\left( \sqrt{\alpha _{2}^{2}-8\alpha _{4}}%
+\alpha _{2}\right) +\frac{12\alpha _{5}}{\sqrt{\alpha _{2}^{2}-8\alpha _{4}}%
}r \\
&&-\frac{5}{8}\left[ \alpha _{2}^{2}-4\alpha _{4}-\frac{\alpha
_{2}^{3}+32\alpha _{6}}{\sqrt{\alpha _{2}^{2}-8\alpha _{4}}}-\frac{72\alpha
_{5}^{2}}{\left( \alpha _{2}^{2}-8\alpha _{4}\right) ^{\frac{3}{2}}}\right]
r^{2}+O(r^{3})\ .  \nonumber
\end{eqnarray}
Since the positivity of the pressure requires $\alpha _{2}>0,$ the density $%
\mu _{-},$ belonging to the negative sign in (\ref{besq}), is necessarily
negative near the center. This shows that if one does not consider
spacetimes with a cosmological constant, the $\beta _{-}$ root of (\ref{ein2}%
) is always unphysical. However, the density $\mu _{+},$ belonging to $\beta
_{+},$ is positive if $\alpha _{4}<0$ and $\alpha _{2}>0.$ In this case the $%
8\alpha _{4}\leq \alpha _{2}^{2}$ condition automatically holds, and hence
the expression under the square root in (\ref{besq}) is always positive. The
dominant energy condition $\mu \geq p$ also holds at the center if $\alpha
_{4}\leq -\frac{2}{9}\alpha _{2}^{2}.$

In a physical situation one expects the pressure and density to have a
maximum at the center. This certainly holds for $p_{+}$ in a neighborhood of
the center, since the coefficient of $r^{2}$ in (\ref{ppex}) is positive
when $\alpha _{2}>0$ and $\alpha _{4}<0.$ If $\alpha _{5}>0$ then $\mu _{+}$
also have a local maximum at the center. In the more realistic case, when $%
\alpha _{5}=0,$ the decreasing nature of $\mu _{+}$ gives a condition on $%
\alpha _{6},$%
\begin{equation}
32\alpha _{6}>-\left( \alpha _{2}^{2}-4\alpha _{4}\right) \sqrt{\alpha
_{2}^{2}-8\alpha _{4}}-\alpha _{2}^{3}\ ,
\end{equation}
which always holds when $\alpha _{6}$ is positive.

Functions $\alpha $ with $\alpha _{5}>0$ seem to be physically realistic,
except that their equation of state for the fluid has some bad properties.
If an equation of state exists in the form $p\equiv p(\mu )$ then 
\begin{equation}
\frac{dp}{dr}=\frac{dp}{d\mu }\frac{d\mu }{dr}
\end{equation}
must hold. However, at the center $\frac{dp}{dr}=0$ but $\frac{d\mu }{dr}$
is nonzero. This shows that $\frac{dp}{d\mu }$ must be zero when $\mu $
takes its central value $\mu _{c},$ which means zero sound speed there.
It would be important to know whether a realistic monotone increasing 
equation of state $p\equiv p(\mu )$ rules out all odd index coefficients 
in the expansion of $\alpha.$

\section{A solution representing a compact fluid \\  sphere}

\label{gensol}The simplest physically realistic polynomial choice for the
generating function $\alpha $ appears to be 
\begin{equation}
\alpha =1+acr^{2}-\frac{1}{8}c^{2}\left( 1-2a^{2}\right) r^{4}\ ,
\end{equation}
where $a$ and $c$ are positive constants satisfying $a^{2}<\frac{1}{2},$ or
rather $a^{2}<\frac{9}{34}$ in order to comply with the dominant energy
condition near the center. For the sake of simplifying the square roots
appearing from (\ref{besq}) we introduce a new radial variable $x$ defined
by 
\begin{equation}
r^{2}=\frac{4\left( \sin x+a\right) }{c\left( 1-2a^{2}\right) }\ .
\end{equation}
Then the variable $x$ is restricted by $x\geq x_{c}=\arcsin \left( -a\right)
.$ The generating function $\alpha $ takes the form 
\begin{equation}
\alpha =\frac{1-2\sin ^{2}x}{1-2a^{2}}\ .
\end{equation}
It still contains the parameter $c$ through 
\begin{equation}
\sin x=\frac{1}{4}c\left( 1-2a^{2}\right) r^{2}-a\ .
\end{equation}
Using that now 
\begin{equation}
r\frac{d}{dr}=\frac{2}{\cos x}(\sin x+a)\frac{d}{dx}
\end{equation}
the function under the square root in (\ref{besq}) becomes 
\begin{equation}
(8\alpha -r\alpha ^{\prime })^{2}-32\alpha (\alpha +1)=\left( 8\cos x\frac{%
\sin x+a}{1-2a^{2}}\right) ^{2}\ .
\end{equation}
The root of (\ref{ein2}) belonging to positive densities takes the simple
form 
\begin{equation}
\beta _{+}=\frac{\sin x+\cos x}{\cos x-a}\ .
\end{equation}
Using equations (\ref{eqB}) and (\ref{eqA}), one of the functions in the
metric is 
\begin{equation}
B_{+}=\frac{(\cos x-a)^{2}\cos (2x)}{(1-2a^{2})\left[ 1+\sin (2x)\right] }\ ,
\end{equation}
while the derivative of the other simplifies to 
\begin{equation}
\frac{d\nu _{+}}{dx}=\frac{\cos x}{2\left( \cos x-a\right) }\ .
\end{equation}
This can be integrated to yield 
\begin{equation}
\nu _{+}=\frac{x}{2}+\frac{a}{\sqrt{1-a^{2}}}\mathrm{arc}\tanh \left( \frac{%
1+a}{\sqrt{1-a^{2}}}\tan \frac{x}{2}\right)
\end{equation}
where an integration constant has been absorbed into the scaling of the time
coordinate $t.$

Using $x$ as the radial coordinate the metric takes the form 
\begin{eqnarray}
ds^{2} &=&-\exp \left[ x+\frac{2a}{\sqrt{1-a^{2}}}\mathrm{arc}\tanh \left( 
\frac{1+a}{\sqrt{1-a^{2}}}\tan \frac{x}{2}\right) \right] dt^{2}  \nonumber \\
&&+\frac{\left[ 1+\sin (2x)\right] \cos ^{2}x}{c\cos (2x)(\sin x+a)(\cos
x-a)^{2}}dx^{2} \\
&&+\frac{4\left( \sin x+a\right) }{c\left( 1-2a^{2}\right) }\left(
d\vartheta ^{2}+\sin ^{2}\vartheta d\varphi ^{2}\right) \ .  \nonumber
\end{eqnarray}
This metric appears to be a new solution. Since the scale of the time
coordinate $t$ is arbitrary, the constant $c$ corresponds to a constant
conformal transformation of the metric. The pressure and density are 
\begin{eqnarray}
8\pi p &=&c\frac{(a-3\sin x)\cos x+(3a-\sin x)\sin x}{4(\sin x+\cos x)} \\
8\pi \mu &=&c\frac{7\sin x-11a+2a\sin ^{2}x-10\sin ^{3}x}{4(\sin x+\cos
x)^{2}}  \nonumber \\
&&+c\frac{3+2a^{2}-4a\sin x-3\sin ^{2}x+6a\sin ^{3}x}{2\cos x(\sin x+\cos
x)^{2}}\ .
\end{eqnarray}
The fluid pressure and density are monotonically decreasing out from a
regular center to a $p=0$ surface. The dominant energy condition $\mu >p$ is
satisfied everywhere for $a<\sqrt{\frac{9}{34}}\approx 0.514,$ and the sound
speed $\frac{dp}{d\mu }$ is positive and less than one for $a<0.184.$
Unfortunately it is very difficult to express the equation of state in $\mu
=\mu (p)$ form, or even in an $f(\mu ,p)=0$ form. The simplest expression
the author could obtain is a complicated polynomial, eighth order in $p,$ $a$
and $c,$ and fourth order in $\mu .$

\section{An infinite gaseous sphere solution}

\label{gensol2}A further simple assumption on the generating function $%
\alpha $ is that it is the ratio of two polynomials of the radial coordinate 
$r.$ Considering the results in Section \ref{secexp}, the lowest degree form
which might give a physically interesting solution is 
\begin{equation}
\alpha =1+\frac{a^{2}r^{2}}{1+br^{2}}=1+a^{2}\left(
r^{2}-br^{4}+b^{2}r^{6}-b^{3}r^{8}\right) +O(r^{10})\ ,
\end{equation}
where $a$ and $b$ are positive constants. It is convenient to introduce a
further constant $c$ defined by 
\begin{equation}
c^{2}=\frac{2}{a^{2}}(b-a^{2})
\end{equation}
and use it in place of the constant $b.$ The assumption that $c$ is real and
non-negative restricts the original constants into the range $b\geq a^{2}.$
In order to simplify the square roots appearing from (\ref{besq}) while
expressing $\beta ,$ we introduce a new radial variable $x$ defined by 
\begin{equation}
r^{2}=\frac{2c-3\sinh x}{a^{2}\left( 2+c^{2}\right) \left( 2\sinh x-c\right) 
}\ .
\end{equation}
Then the center is at $x_{c}=\mathrm{arc}\sinh \frac{2c}{3},$ spatial
infinity is at $x_{\infty }=\mathrm{arc}\sinh \frac{c}{2},$ and the new
variable is restricted by $0<x_{\infty }\leq x\leq x_{c}.$ The $b\leq a^{2}$
case could be treated in a similar way introducing sinus functions instead
of sinus hyperbolics, but would lead to solutions which fail to satisfy the
dominant energy condition $\mu \geq p$ at infinity. The generating function $%
\alpha $ takes the form 
\begin{equation}
\alpha =\frac{4c+\left( c^{2}-4\right) \sinh x}{\left( 2+c^{2}\right) \sinh x%
}\ .
\end{equation}
The function under the square root in (\ref{besq}) becomes a square 
\begin{equation}
(8\alpha -r\alpha ^{\prime })^{2}-32\alpha (\alpha +1)=\left( \frac{8c\left(
2c-3\sinh x\right) \cosh x}{\left( 2+c^{2}\right) \sinh ^{2}x}\right) ^{2}\ .
\end{equation}
The root of (\ref{ein2}) belonging to positive densities takes the form 
\begin{equation}
\beta _{+}=\frac{\left( c\coth \frac{x}{2}-2\right) \cosh \frac{x}{2}}{\cosh 
\frac{x}{2}+c\sinh \frac{x}{2}}\ .
\end{equation}
Using equation (\ref{eqB}), one of the functions in the metric is 
\begin{equation}
B_{+}=\frac{\left[ 4c+\left( c^{2}-4\right) \sinh x\right] \left( c\tanh 
\frac{x}{2}+1\right) ^{2}}{(2+c^{2})\left( c\coth \frac{x}{2}-2\right)
^{2}\sinh x}\ ,  \label{bb2}
\end{equation}
while, from (\ref{eqA}), the derivative of the other function becomes 
\begin{equation}
\frac{d\nu _{+}}{dx}=\frac{c\cosh x}{4\sinh \frac{x}{2}\left( \cosh \frac{x}{%
2}+c\sinh \frac{x}{2}\right) \left( c-2\sinh x\right) }\ .
\end{equation}
This can be integrated to obtain 
\begin{eqnarray}
\nu _{+} &=&\frac{1}{2}\ln \sinh \frac{x}{2}+\frac{1}{2\left( 3+c^{2}\right) 
}\left[ 2\sqrt{4+c^{2}}\mathrm{arc}\tanh \left( \frac{2+c\tanh \frac{x}{2}}{%
\sqrt{4+c^{2}}}\right) \right.  \nonumber \\
&&\left. -\left( 1+c^{2}\right) \ln \left( \cosh \frac{x}{2}+c\sinh \frac{x}{%
2}\right) -\ln \left( 2\sinh x-c\right) \right] \ .  \label{nup2}
\end{eqnarray}

Using $x$ as the radial coordinate the metric takes the form 
\begin{eqnarray}
ds^{2} &=&-\exp \left( 2\nu _{+}\right) dt^{2}+\frac{2c-3\sinh x}{%
a^{2}\left( 2+c^{2}\right) \left( 2\sinh x-c\right) }\left( d\vartheta
^{2}+\sin ^{2}\vartheta d\varphi ^{2}\right)   \nonumber \\
&&+\frac{c^{2}\cosh ^{2}x}{4a^{2}\left( 2+c^{2}\right) \left( 2c-3\sinh
x\right) \left( 2\sinh x-c\right) ^{3}B_{+}}dx^{2}\ ,
\end{eqnarray}
where $\nu _{+}$ is determined by equation (\ref{nup2}) and $B_{+}$ by (\ref
{bb2}). This second metric also appears to be a new solution. The constant $a
$ corresponds to a constant conformal transformation of the metric. The
pressure and density are 
\begin{eqnarray}
8\pi p &=&a^{2}\frac{(2\sinh x-c)(2-2c^{2}+2\cosh x+5c\sinh x)}{4(c\cosh 
\frac{x}{2}-2\sinh \frac{x}{2})\cosh ^{3}\frac{x}{2}} \ , \\
8\pi \mu  &=&\frac{a^{2}(2\sinh x-c)}{32\left( c\cosh \frac{x}{2}-2\sinh 
\frac{x}{2}\right) ^{2}\cosh ^{4}\frac{x}{2}\cosh x}\left[ 3\left(
7c^{2}-12\right) \sinh (3x)\right.   \nonumber \\
&&+6c\left( 3c^{2}-4\right) \cosh (3x)+2\left( 12+61c^{2}-8c^{4}\right)
\sinh (2x)  \nonumber \\
&&+4c\left( 62-13c^{2}\right) \cosh (2x)+\left( 156-475c^{2}+32c^{4}\right)
\sinh x \\
&&\left. +2c\left( 4-53c^{2}\right) \cosh x+12c\left( 13c^{2}-22\right) 
\right]  \ . \nonumber
\end{eqnarray}
As indicated by numerical plots of the quantities, the fluid pressure and
density are monotonically decreasing to zero, out from a regular center to
infinity. For any choice of the parameters $a$ and $c,$ the dominant energy
condition $\mu >p$ is satisfied, and the sound speed $\frac{dp}{d\mu }$ is
positive and less than one. Eliminating the variable $x,$ one can obtain a
complicated polynomial equation of state in the form $f(\mu ,p)=0,$ eighth
order in $p,$ fourth order in $\mu ,$ and twelfth order in $a$ and $c.$ Near
spatial infinity, in the small $\mu $ and $p$ limit, the equation of state
is approximately linear,
\begin{equation}
\frac{p}{\mu }=\frac{6+4\sqrt{4+c^{2}}}{2\left( 4c^{2}+7\right) }<1\ .
\end{equation}

\section{Conclusions}

An algorithm has been given, which can be used to generate physically
realistic density and pressure distributions from a generating function 
$\alpha$ without calculating integrals. 
Any function $\alpha$ of which the first few expansion coefficients
satisfy the simple conditions stated in Section \ref{secexp} generate 
spacetimes which are physically well behaving at least in a neighborhood 
of the center.
The resulting pressure and density distributions can contain arbitrarily
many parameters, for example by choosing $\alpha$ to be a high order 
polynomial.
Unfortunately, the actual forms of $\mu $ and $p$ can be quite complicated 
because of the square roots appearing. 
This makes the task of putting the resulting equation of state in
a closed form very difficult.
Unfortunately, a prescribed equation of state would yield a very complicated
differential equation on the generating function.
However, by trying many different functional forms for $\alpha$,
hopefully one could find configurations with simple and physical
equations of states.

The metric function $\nu$ only appears in a differentiated form in the
field equation and in the density and pressure expressions. 
Correspondingly, only its derivative $\nu^{\prime}$ can be given as an 
algebraic expression of $\alpha$ and its first and second derivatives. 
Because of this, in some sense, one could consider a generated solution
as an "exact solution" even if only the derivative of $\nu$ can be given in
a closed form.
However, for some simpler choices of $\alpha ,$ the integral determining 
$\nu$ can be calculated.
This has been the case for the two example solution solutions presented
in this paper.

It would be important to find out whether the generating formalism could be
made simpler, or the obtainable exact solutions and equations of states
could become more physical, for example, by choosing a functional expression
$f(\alpha)$ of the generating function instead of $\alpha ,$ or by using a
different radial coordinate in place of the area coordinate $r.$

\section*{Acknowledgments}

This work was supported by OTKA grant T022533 and by the Japan Society for
the Promotion of Science.

\end{document}